\documentclass[twocolumn]{aastex631}

\usepackage{amsmath}
\usepackage{graphicx}
\usepackage{longtable}

\definecolor{myblue}{RGB}{0, 190, 245} 
\definecolor{darkblue}{RGB}{0, 0, 200} 
\definecolor{mygreen}{RGB}{0, 180, 0}

\newcommand{\kms}{km s$^{-1}$ Mpc$^{-1}$}

\defcitealias{Riess2019}{R19}

\hypersetup{colorlinks = true, 
citecolor = {blue},  
linkcolor={darkblue}, 
urlcolor={darkblue}}

\shorttitle{A New Anchor for the SH0ES Distance Ladder}
\shortauthors{Breuval L., et al.}

\begin{document}


\title{Small Magellanic Cloud Cepheids Observed with the Hubble Space Telescope \\ Provide a New Anchor for the SH0ES Distance Ladder}

\author[0000-0003-3889-7709]{Louise Breuval}
\affiliation{Department of Physics and Astronomy, Johns Hopkins University, Baltimore, MD 21218, USA}
\email{lbreuva1@jhu.edu}

\author[0000-0002-6124-1196]{Adam G. Riess} 
\affiliation{Department of Physics and Astronomy, Johns Hopkins University, Baltimore, MD 21218, USA}
\affiliation{Space Telescope Science Institute, 3700 San Martin Drive, Baltimore, MD 21218, USA}

\author{Stefano Casertano} 
\affiliation{Space Telescope Science Institute, 3700 San Martin Drive, Baltimore, MD 21218, USA}

\author[0000-0001-9420-6525]{Wenlong Yuan}
\affiliation{Department of Physics and Astronomy, Johns Hopkins University, Baltimore, MD 21218, USA}

\author[0000-0002-1775-4859]{Lucas M.~Macri}
\affiliation{NSF NOIRLab, 950 N. Cherry Ave, Tucson AZ 85719, USA}

\author[0000-0002-5527-6317]{Martino Romaniello}
\affiliation{European Southern Observatory, Karl-Schwarzschild-Strasse 2, 85478 Garching bei München, Germany}

\author[0000-0002-8342-3804]{Yukei S. Murakami}
\affiliation{Department of Physics and Astronomy, Johns Hopkins University, Baltimore, MD 21218, USA}

\author[0000-0002-4934-5849]{Daniel Scolnic}
\affiliation{Department of Physics, Duke University, Durham, NC 27708, USA}

\author[0000-0002-5259-2314]{Gagandeep S. Anand} 
\affiliation{Space Telescope Science Institute, 3700 San Martin Drive, Baltimore, MD 21218, USA}

\author[0000-0002-7777-0842]{Igor Soszy\'nski}
\affiliation{Astronomical Observatory, University of Warsaw, Al. Ujazdowskie 4, 00-478 Warszawa, Poland}

\begin{abstract}
We present phase-corrected photometric measurements of 88 Cepheid variables in the core of the Small Magellanic Cloud (SMC), the first sample obtained with the \textit{Hubble} Space Telescope's (\textit{HST}) Wide Field Camera 3, in the same homogeneous photometric system as past measurements of all Cepheids on the SH0ES distance ladder. We limit the sample to the inner core and model the geometry to reduce errors in prior studies due to the nontrivial depth of this cloud. Without crowding present in ground-based studies, we obtain an unprecedentedly low dispersion of 0.102~mag for a period-luminosity (P-L) relation in the SMC, approaching the width of the Cepheid instability strip. The new geometric distance to 15 late-type detached eclipsing binaries in the SMC offers a rare opportunity to improve the foundation of the distance ladder, increasing the number of calibrating galaxies from three to four. With the SMC as the only anchor, we find H$_0\!=\!74.1 \pm 2.1$ \kms. Combining these four geometric distances with our \textit{HST} photometry of SMC Cepheids, we obtain H$_0\!=\!73.17 \pm 0.86$ \kms. By including the SMC in the distance ladder, we also double the range where the metallicity ([Fe/H]) dependence of the Cepheid P-L relation can be calibrated, and we find $\gamma = -0.234 \pm 0.052$~mag~dex$^{-1}$. Our local measurement of H$_0$ based on Cepheids and Type Ia supernovae shows a 5.8$\sigma$ tension with the value inferred from the cosmic microwave background assuming a Lambda cold dark matter ($\Lambda$CDM) cosmology, reinforcing the possibility of physics beyond $\Lambda$CDM.   \\ 
\end{abstract}

\section{Introduction} 

The tension between the local measurement of the Hubble constant (H$_0$) based on distances and redshifts \citep[e.g., H$_0$ = 73.0 $\pm$ 1.0 $\rm km \, s^{-1} \, Mpc^{-1}$][]{Riess2022a} and its value inferred from the Lambda cold dark matter ($\Lambda$CDM) model calibrated with cosmic microwave background data in the early Universe \citep[e.g., H$_0$ = 67.4 $\pm$ 0.5 $\rm km \, s^{-1} \, Mpc^{-1}$][]{Planck2020} has reached a $5\sigma$ significance.  Extensive, recent reviews of the measurements are provided by \cite{DiValentino2021} and \cite{Verde2023}. This intriguing discrepancy provides growing hints of new physics beyond $\Lambda$CDM, which might include exotic dark energy, new relativistic particles, neutrino interactions, or a small curvature. Continued efforts to test $\Lambda$CDM in the late and early Universe to identify departures and tensions are warranted.  While this tension is well documented, increasing the number of calibrations or anchors of the local distance measurements is of great importance in order to better characterize and quantify the size of the tension.

\begin{figure*}[t!]
\includegraphics[height=7.4cm]{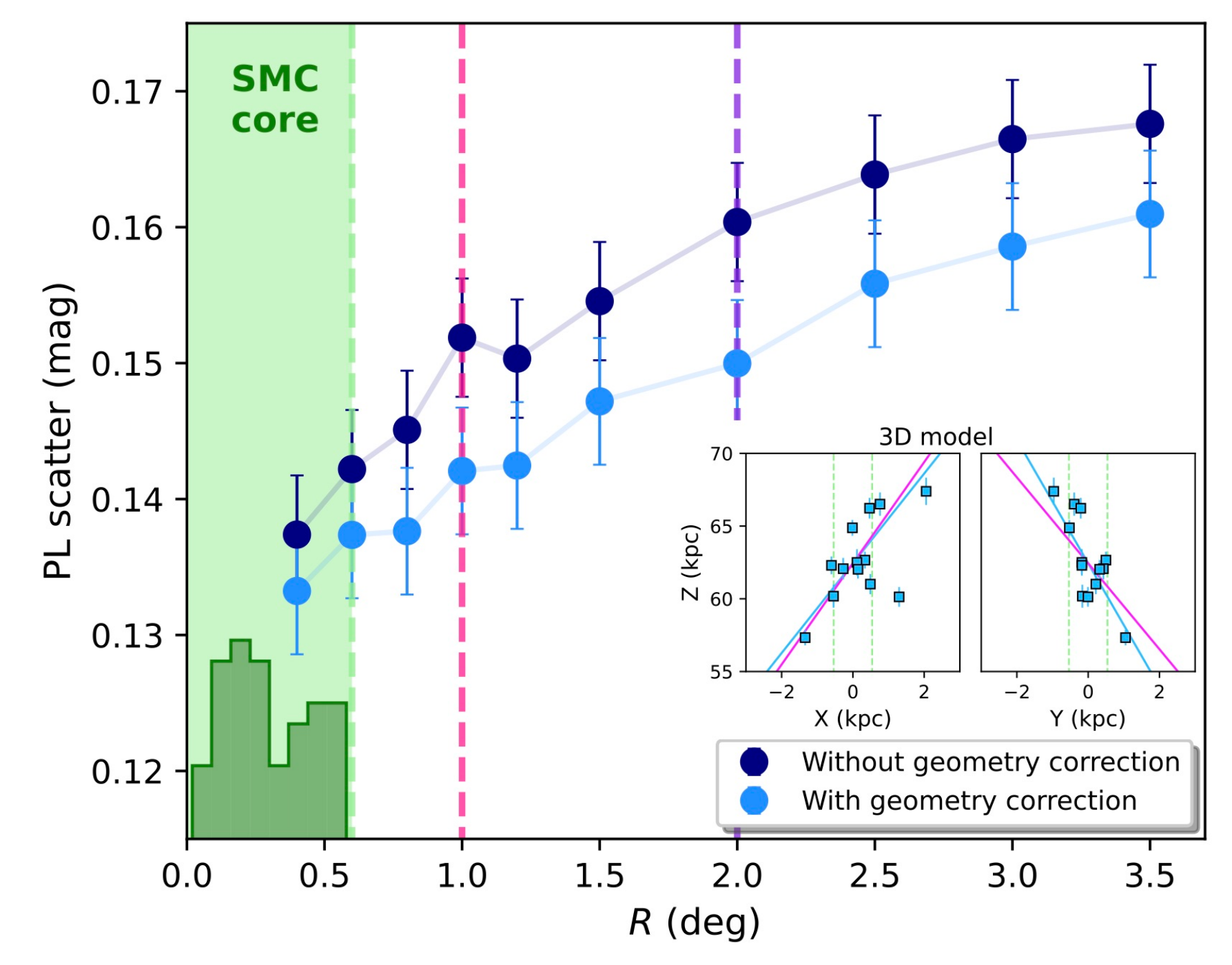} ~
\includegraphics[height=7.4cm]{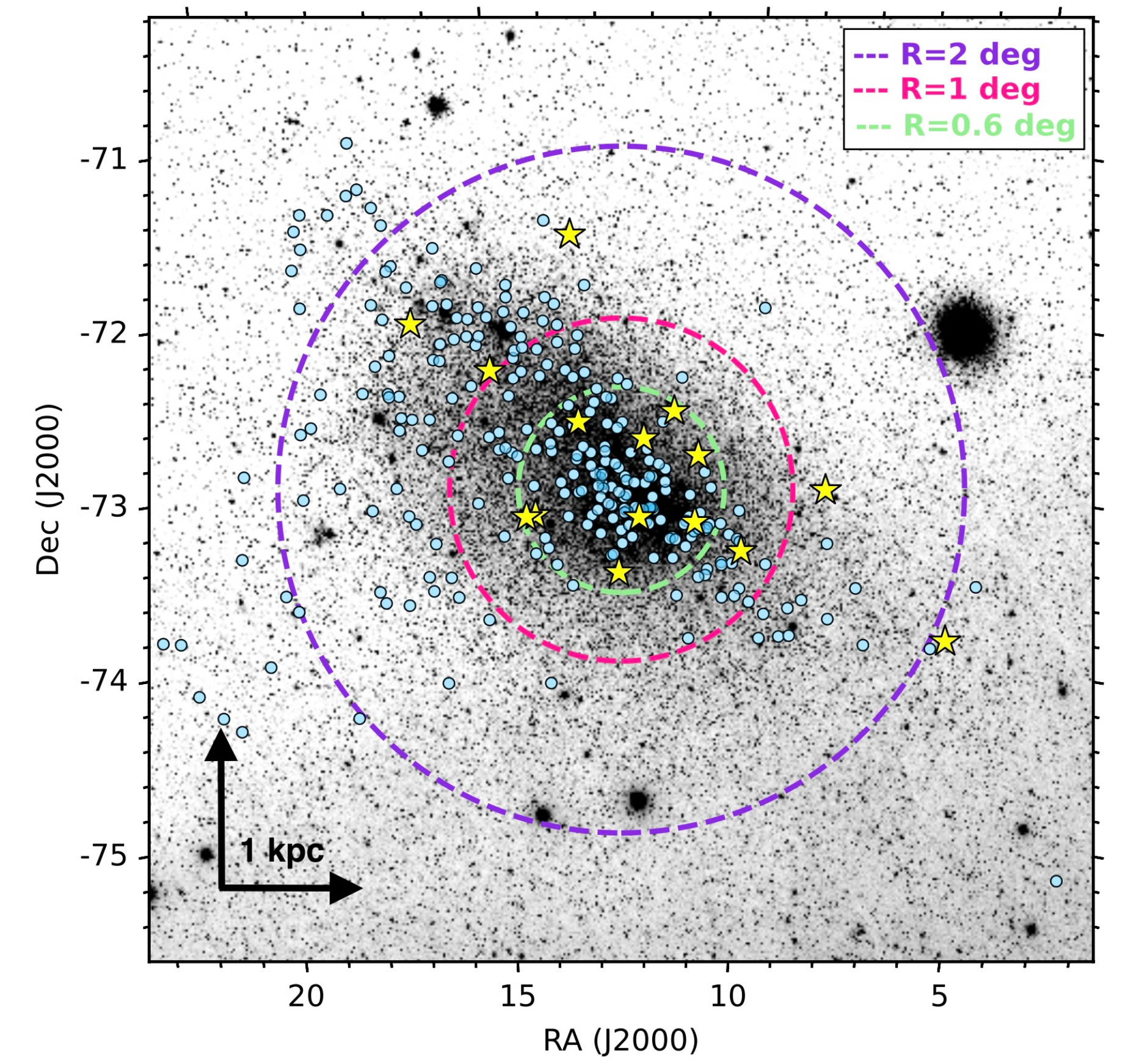}
\caption{{\bf (Left)}: P--L scatter in the NIR $W_{JK}$ Wesenheit index for different Cepheid subsamples located at increasing distances from the SMC center. For this test, Cepheid photometry was taken from the VISTA near-infrared YJKs survey of the Magellanic Clouds system \citep[VMC;][]{Cioni2011} and the geometric corrections were derived using the planar geometry by \cite{Graczyk2020}. The green histogram in the bottom left corner shows the distribution of our Cepheid sample. The small inset on the right shows the position of DEBs from \cite{Graczyk2020} on the XZ and YZ planes, with the blue and pink lines showing the model fits from DEBs and from our Cepheid sample, respectively. {\bf (Right)}: Map of SMC Cepheids from \cite{Ripepi2017} in light blue with the regions corresponding to radii of 2 deg, 1 deg and 0.6 deg. Eclipsing binaries from \cite{Graczyk2020} are shown in yellow. \\
\label{fig:PL_scatter_vs_radius}}
\end{figure*}

Cepheid variables have been the ``gold standard'' of primary distance indicators for over a century.  They are well-understood pulsating stars \citep{Eddington1917, Zhevakin1963} and are easy to identify thanks to their large-amplitude light curves at optical wavelengths. The Cepheids in the Small Magellanic Cloud (SMC) in particular played a foundational role in the recognition of their use as distance indicators and subsequent application to the discoveries of extragalactic nature of galaxies and the expansion of the Universe \citep{Hubble1929}. In the early 1900s, Henrietta Leavitt \citep{Leavitt1908} discovered a remarkable phenomenon by examining photographic plates of the SMC: the brightness of Cepheids follows a linear relationship with the logarithm of their period, with the brightest Cepheids having the longest periods. This law is now called the Leavitt law \citep{Leavitt1912} or simply the period-luminosity (P--L) relation. 

Since this discovery, SMC Cepheids have been used extensively to map the structure of this galaxy \citep{Scowcroft2016, Jacyszyn2016, Ripepi2017, Deb2019}, to investigate possible non-linearities in the P--L relation \citep{Sandage2009}, to study the chemical composition of these pulsating stars \citep{Lemasle2017}, or to calibrate the Cepheid metallicity dependence \citep{Wielgorski2017, Gieren2018, Breuval2021, Breuval2022}. SMC Cepheids have been observed by large surveys, and complete light curves have been obtained at multiple wavelengths \citep{Soszynski2015, Scowcroft2016, Ripepi2017, Ripepi2023}. As a nearby dwarf galaxy and a direct neighbor of the Milky Way (MW), the SMC is a great laboratory for many astrophysical studies. However, the downside of this galaxy, in particular for distance measurements, is its elongated shape along the line of sight, which produces a detectable range of 5-10\% in the distances to its stars. The consequence is that SMC Cepheids show a higher scatter than their Large Magellanic Cloud (LMC) cousins. However, it is possible to reduce the variation in depth by limiting studies to the SMC core (rather than the far-flung streams and condensations that surround it) and by the application of a geometric model recently derived from a study of the detached eclipising binaries (DEBs) near the core, as shown in Fig.~\ref{fig:PL_scatter_vs_radius}; the Cepheid dispersion is reduced if a geometric model fit to the DEBs is applied (light blue vs.~dark blue points in Fig.~\ref{fig:PL_scatter_vs_radius}). The geometric correction assumed here is based on a planar geometry, such as the one from \cite{Graczyk2020}. In the past, these corrections have often been neglected, resulting in a large scatter and even a bias in the P--L intercept and in the metallicity dependence \citep{Wielgorski2017, Owens2022}; see also the discussion in \cite{Breuval2022}.  By limiting to the core and accounting for the depth, ground-based near-infrared (NIR) samples produce a P--L scatter of $\sim$ 0.13~mag, approaching the state of the art.  

\begin{figure*}[t!]
\includegraphics[width=9.0cm]{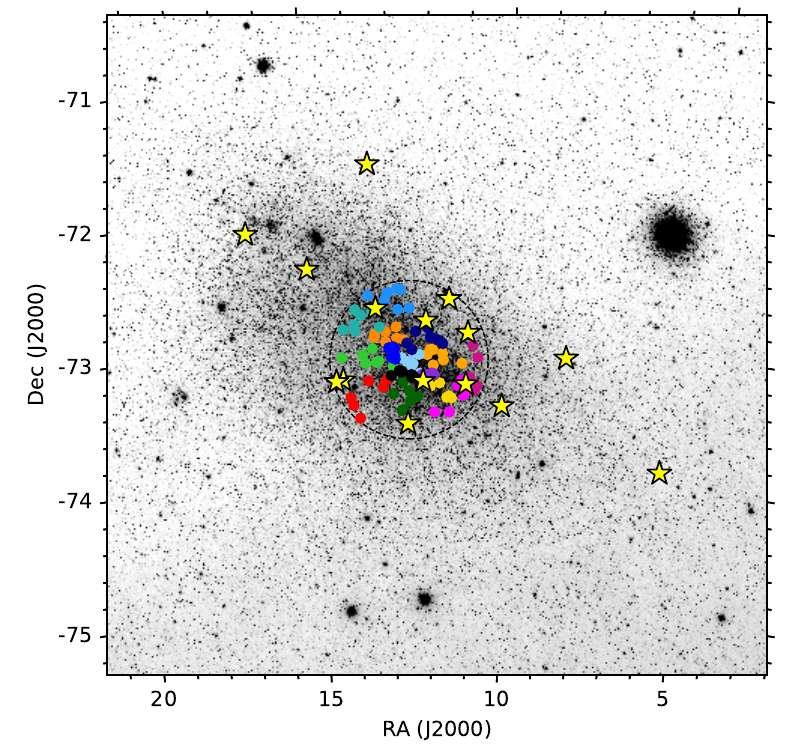}
\includegraphics[width=9.0cm]{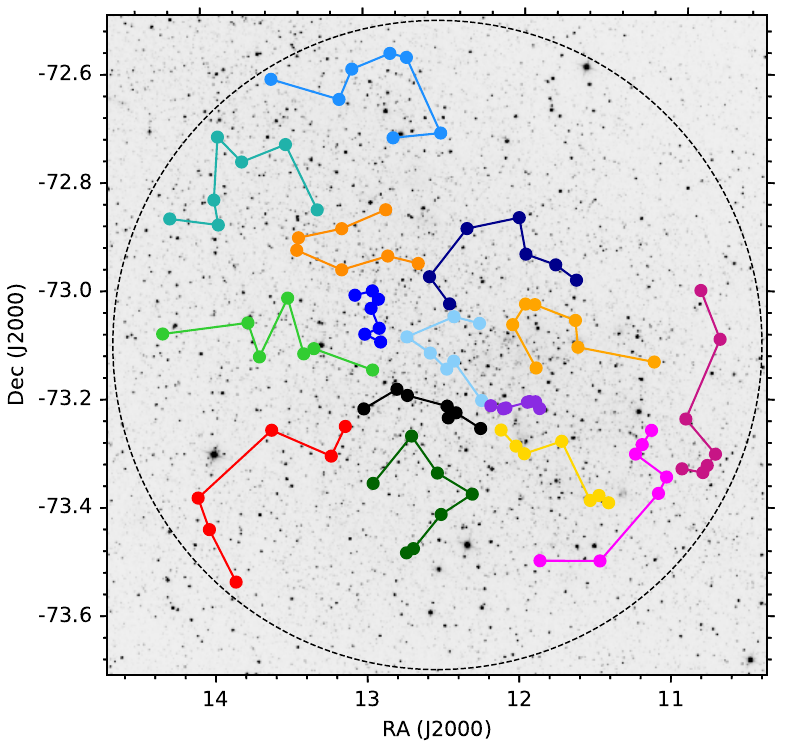}
\caption{{\bf (Left)}: The dashed circle shows the core region of the SMC ($R=0.6$ deg). DEBs from \cite{Graczyk2020} are shown in yellow and the colored dots are the Cepheids from the present study. {\bf (Right)}: Zoom in of the SMC core region. The 15 Cepheid sequences (one per \textit{HST} orbit) are shown in color. \\
\label{fig:SMC_map}}
\vspace{0.5cm}
\end{figure*}

The new geometric distance to the SMC by \cite{Graczyk2020}, based on late-type DEBs, provides a unique opportunity to improve the distance ladder by increasing the number of anchor galaxies from three to four. One of the strengths of the SH0ES distance ladder is in the consistency of the photometric measurements across all of the first and second rungs: \textit{all} Cepheid photometry is obtained in a single homogeneous photometric system with the \textit{Hubble} Space Telescope (\textit{HST}) Wide Field Camera 3 (WFC3), the only instrument capable of reaching Cepheids in Type Ia supernovae (SNe~Ia) hosts as well as observing nearby Cepheids in the Milky Way, therefore nulling zero-point errors. Indeed, combining ground-based and \textit{HST} photometry would likely produce a $1.4-1.8$\% systematic error in the measured distances simply due to bandpass differences \citep{Riess2019}. Also, due to the density of stars in the SMC core, typical ground-based samples with seeing worse than an arcsecond suffer crowding. Because of the lack of \textit{HST} photometry in the same system, the SMC could not be previously included as an anchor of the SH0ES distance ladder for calibrating Cepheids \citep{Riess2022a}. In order to make full use of the new geometric distance to the SMC, we observed a sample of Cepheids in the core of the SMC to mitigate the depth effects, using HST resolution instead of ground-based photometry to avoid crowding, and in the same photometric system (HST/WFC3) used for the SH0ES distance ladder to ensure consistency and negate photometric zero-points. As we show, these further reduce the scatter to a best-seen 0.10~mag while putting all Cepheids on the same photometric system as the rest of the distance ladder.

We describe our observations in \S\ref{sec:observations} and the Cepheid photometry and phase corrections in \S\ref{sec:photometry}. The P--L relations, as well as a discussion of the effects of metallicity and of the SMC geometry, are presented in \S\ref{sec:PLR}. The Hubble constant is derived in \S\ref{sec:H0} from our \textit{HST} photometry and the previous SH0ES data. Finally, we discuss the results and prospects for H$_0$ in \S\ref{sec:discussion}.  \\


\section{Observations}
\label{sec:observations}

\subsection{Sample selection}

The data used in this paper were obtained as part of the \textit{HST} Cycle 30 program GO-17097 (PI: A.~Riess) between 2023 June 23 and October 4. We selected our sample based on the OGLE-IV catalog of Magellanic Cloud Cepheids \citep{Soszynski2015}. To decrease depth effects and to reduce the P--L scatter (Fig.~\ref{fig:PL_scatter_vs_radius}), we excluded Cepheids beyond 0.6~deg from the SMC center \citep[$\alpha = 12.54 \, \rm deg$, $\delta = -73.11 \, \rm deg$, from][]{Ripepi2017}. This region is small enough to mitigate the depth effects and contains enough Cepheids to populate the P--L relation. We discuss the effects of the SMC geometry in \S\ref{sec:geocorr}. Finally, we selected Cepheids in the range $0.7<\log P < 2.0$, with $P$ the pulsation period in days, similar to \cite{Riess2019} in the LMC, to avoid contamination from shorter-period overtone Cepheids ($\sim$46\% of all SMC Cepheids) and possible non-linearities in the P--L relation at shorter periods \citep{Bhardwaj2016b, Bhardwaj2020, Soszynski2024}. \\

\begin{figure}[t!]
\includegraphics[width=8.2cm]{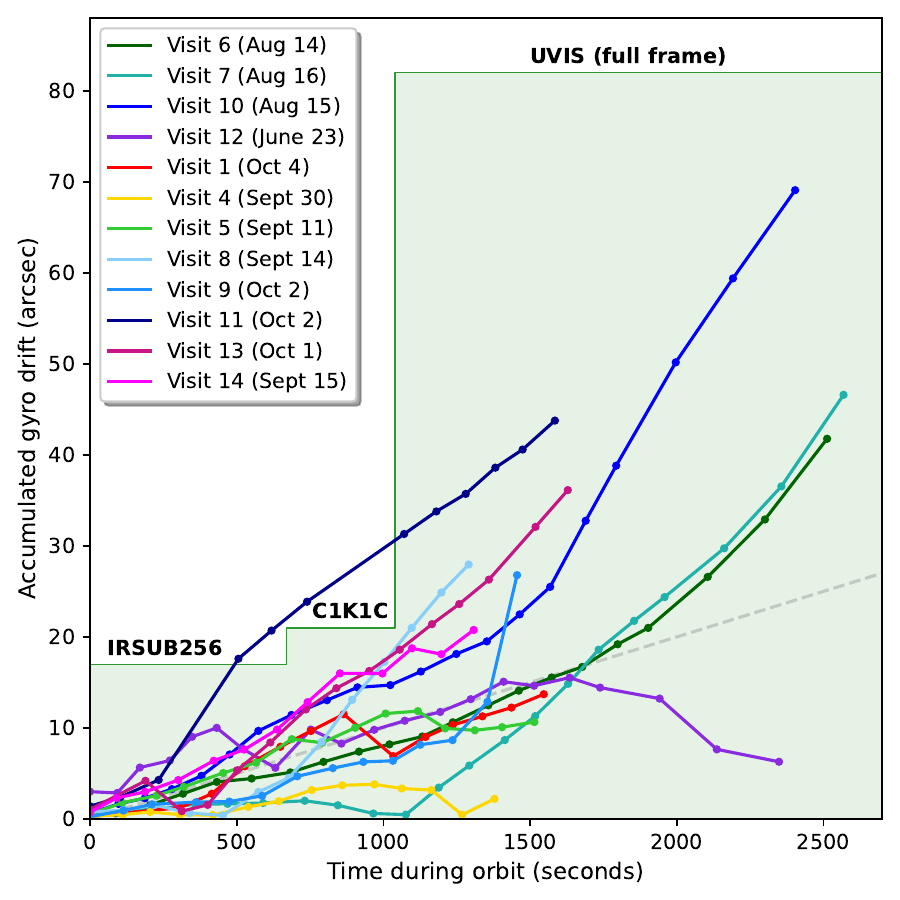}
\caption{Accumulated drift of the pointing due to the gyroscopic control in the ``DASH" mode of \textit{HST}. The green region shows the size of the image (with the subarray names) and the dashed grey line is the expected drift of 0.01\arcsec/sec (STScI handbook). Visits 2, 3 and 15 were observed under FGS control and therefore are not shown in this figure. The failed visits observed in the first attempt (with drifts up to 200\arcsec) are not represented in this figure.
\label{fig:gyro_drift}}
\end{figure}

\subsection{The DASH mode}

Because the SMC is a very nearby galaxy, the mean separation between Cepheids is large ($\sim 12\arcmin$) compared to the WFC3 field of view ($\sim 2.5\arcmin$).  It is therefore inefficient to observe SMC Cepheids individually with {\textit{HST}} using normal pointing procedures, which require $\sim6-7$ minutes for a guide-star lock every time a new position is acquired.   However, the rapid-exposure observing mode of \textit{HST}, called ``Drift And SHift" \citep[or ``DASH";][]{Momcheva2017}, available since Cycle 24, makes the observations much more efficient than the classic procedure by slewing the telescope a few arcminutes between Cepheid pairs under gyroscope control, allowing us to observe multiple short-exposure targets in a single orbit after a single guide star acquisition. This technique was successfully adopted by \cite{Riess2019}, hereafter \citetalias{Riess2019}, to observe 70 LMC Cepheids (GO-14648, GO-15145), which enabled the use of this galaxy as a primary anchor of the distance ladder. In the following, we adopt the approach from \citetalias{Riess2019} to ensure consistency between Cepheid measurements in both galaxies and to negate systematic uncertainties in the photometry.

We selected 15 sequences of 6 or 7 Cepheids such that the slewing of the telescope was minimized, as shown in the right panel of Fig.~\ref{fig:SMC_map}.  The length of each sequence was chosen so that each could be observed during one \textit {HST} visibility period, typically  $\sim2.5~$ksec. For a given sequence, we first observed each Cepheid in the NIR $F160W$ filter, then flipped the WFC3 mirror mechanism and reversed the path to observe them in UVIS filters $F555W$ and $F814W$. To avoid the loss of efficiency that would be caused by a memory dump, we used subarrays instead of full frame images. We started with 33\arcsec\  subarrays in the NIR (256×256 pixels, IRSUB256), then 40\arcsec\  subarrays in the first optical exposures (1K×1K pixels, C1K1C) and larger 80\arcsec\  subarrays (2K×2K pixels, 2K2C) for the last optical images. This successive expansion of the array size was intended to keep the target in the field of view even with a gyro drift rate of up to twice the expected value.

The first observation (visit 12) was obtained successfully on 2023 June 23 with an accumulated drift of 15\arcsec\  by the end of the visit, keeping all Cepheids well within the subarrays and providing images of seven targets in three filters. The following visits, observed between 2023 August 10-16, were severely affected by a period of erratic performance of gyro 3, causing large drifts up to ten times larger than expected, placing most Cepheids outside the frame. 

After a number of missed targets, we adopted a new strategy and redesigned subsequent and repeat observations to complete the exposures in the missing filters and to allow for a greater drift. For the visits that completely failed (gyro drift larger than the array size in the first two exposures), we repeated the observations with full frame images (BIN=2) for the last third of the orbit to accommodate larger drifts. These full frame images require more memory, thus we were only be able to fit five targets per orbit. For the visits that had partial success (some Cepheids were observed but not in all three filters), we repeated only the missed filters under Fine Guidance Sensor (FGS) control, independent of the gyro behavior. In this configuration, we were able to fit four targets per orbit. In some cases, the images were large enough to observe two very nearby Cepheids in the same frame. The 7 rescoped visits were successfully executed between 2023 September 11 and October 4. Our final sample comprises 88 Cepheids (out of the 104 proposed), representing only a net 9\% loss in the statistical power of the sample. 


\begin{table*}[t!]
\caption{Observations of SMC Cepheids}
\begin{tabular}{c c c c c c c c}
\hline
\hline
Cepheid & Frame & Filter & MJD & Exposure & Array & X$_{\rm Cepheid}$ & Y$_{\rm Cepheid}$ \\
 &  &  & \scriptsize{(days)} & \scriptsize{time (s)} &  & \multicolumn{2}{c}{\scriptsize{(pixels)}} \\
\hline
OGLE-0518 & iev9tuh3q & F160W & 60218.482 & 2.50 & IRSUB256 & 138.4 & 129.2  \\
OGLE-0518 & iev9tuhcq & F555W & 60218.490 & 2.50 & UVIS1 & 843.2 & 680.8  \\
OGLE-0518 & iev9tuhdq & F814W & 60218.492 & 2.50 & UVIS1 & 817.8 & 706.1  \\
OGLE-0524 & iev9tuh4q & F160W & 60218.483 & 2.50 & IRSUB256 & 126.2 & 142.0  \\
OGLE-0524 & iev9tuhbq & F555W & 60218.489 & 2.50 & UVIS2-C1K1C-SUB & 195.5 & 796.1  \\
OGLE-0524 & iev9tuhaq & F814W & 60218.488 & 2.50 & UVIS2-C1K1C-SUB & 231.3 & 761.5  \\
OGLE-0524 & iev9tuh4q & F160W & 60218.483 & 2.50 & IRSUB256 & 126.2 & 142.0  \\
OGLE-0524 & iev9tuhbq & F555W & 60218.489 & 2.50 & UVIS2-C1K1C-SUB & 195.5 & 796.1  \\
OGLE-0524 & iev9tuhaq & F814W & 60218.488 & 2.50 & UVIS2-C1K1C-SUB & 231.3 & 761.5  \\
OGLE-0541 & iev9tuh1q & F160W & 60218.478 & 2.50 & IRSUB256 & 130.7 & 141.8  \\
OGLE-0541 & iev9tuhgq & F555W & 60218.496 & 2.50 & UVIS1 & 720.8 & 808.3  \\
OGLE-0541 & iev9tuhhq & F814W & 60218.497 & 2.50 & UVIS1 & 683.9 & 845.0  \\
\hline
\end{tabular}
{\flushleft \textbf{Notes:} $^{(*)}$ Indicates a drift rate larger than 0.08\arcsec during the exposure necessitating a larger photometry aperture.} \\ \par
{\center (This table is available in its entirety in machine-readable form.) \\ }
 ~ \\
\label{table:observations}
\end{table*}

The accumulated drift caused by the gyroscope guiding is plotted as a function of time during each orbit in Fig.~\ref{fig:gyro_drift}. With a typical accumulated drift of 30\arcsec\  at the end of a visit (2.5~ksec), we expect the Cepheid flux to be smeared by 0.7 pixel (0.03\arcsec) at most during a 2.5-second exposure on the UVIS detector, which we can accomodate with a large photometric aperture. Due to the gyroscope control, Cepheids are often located far from the center of the images and their coordinates are incorrect (in this mode, \textit{HST} cannot use the astrometry of a guide star to establish the pointing coordinates). We therefore used the positions of the brightest stars from the \textit{Gaia} EDR3 catalog \citep{GaiaEDR3_contents} that we find in our images to identify the Cepheid positions. For some of the observations with the largest drifts, direct position matching was not sufficient; in such cases, we used a histogram-based algorithm to match source positions to the \textit{Gaia} catalog. This algorithm is robust to the presence of outliers, e.g., due to cosmic rays, and works up to offsets well over $ 10\arcmin$.  This method allowed us to estimate drifts from the intended pointing that in some cases exceeded several arcminutes.  However, none of the successful observations required this approach.  The observations are described in Table~\ref{table:observations}.  \\

\begin{figure*}[t!]
\centering
\includegraphics[width=18.0cm]{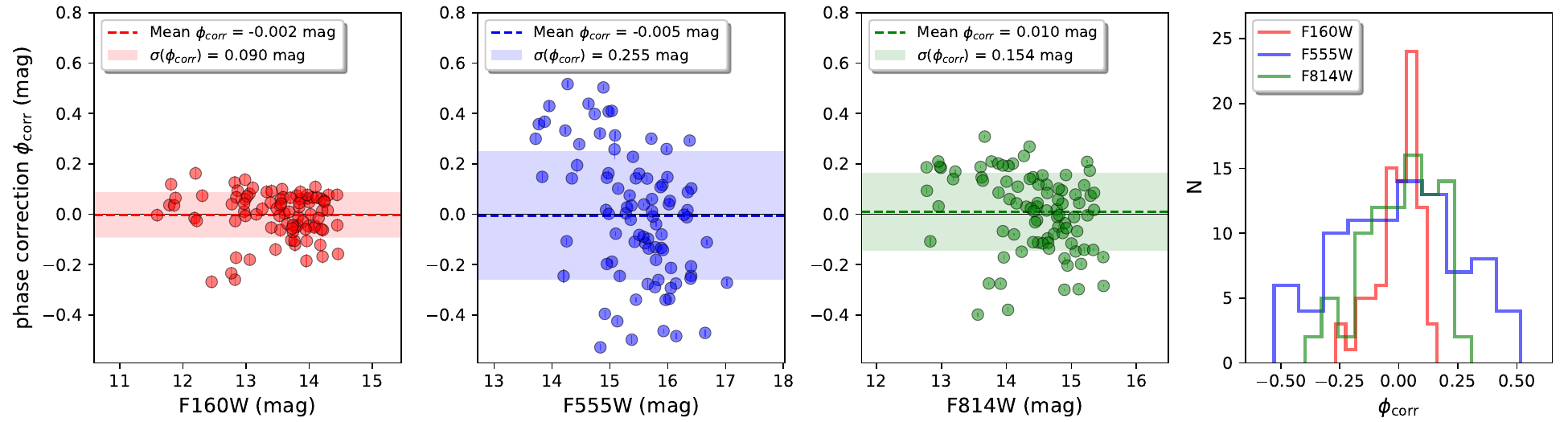}
\caption{Phase corrections derived as the difference between the random epoch magnitude and the intensity averaged mean magnitude in three filters. \\ 
\label{fig:phase_corrections}}
\end{figure*}

\section{SMC Cepheid photometry} 
\label{sec:photometry}

\subsection{Aperture photometry} 

In this section we describe the procedure to measure the flux associated to each Cepheid in the three filters. To accommodate for possible variations in the PSF due to gyro drifts, we perform aperture photometry following the approach described in \citetalias{Riess2019}. We measured the photometry in ``FLC" files in the optical, which are corrected for the charge transfer efficiency (CTE), and ``FLT" files in the NIR. The images were multiplied by pixel area maps to account for the different on-sky pixel size across the field of view. In the case of UVIS full frame images (2$\times$2 binned), we constructed and applied a 2$\times$2 binned pixel area map.

For most images, we use a photometric aperture with radius of 3 pixels, small enough to limit the effect of possible contamination from cosmic rays or nearby stars; after inspecting the images, we found no evidence of such contaminations. We estimate and subtract the background using a sky annulus with radius of 30 to 50 pixels around the Cepheids \citepalias[same as in][]{Riess2019}. For the UVIS full-frame binned images, we adopt the equivalent: an aperture radius of 1.5 pixels and a sky annulus of 15 to 25 pixels for the background. The Cepheid magnitude $m_{\lambda}$ is derived as follows:
\begin{equation}
m_{\lambda} = m_{\lambda}^0 + (zp -25) - \rm ap_{corr}
\label{eq:photometry}
\end{equation}
where $m_{\lambda}^0$ is the measured magnitude, $zp$ is the magnitude of a star which produces a total count rate of $1 \, \rm e^- s^{-1}$,and ap$_{\rm corr}$ is the aperture correction from a 3 pixel radius to infinity. We adopted the zero-points from \citetalias{Riess2019} and provided by STScI for each UVIS CCD and each filter (Vega system): 24.711~mag in $F160W$, 25.741~mag and 24.603~mag in $F555W$ and $F814W$ respectively in chip~1, and 25.727~mag and 24.581~mag in $F555W$ and $F814W$ respectively in chip~2. We followed \citetalias{Riess2019} and adopted aperture corrections of 0.200, 0.183 and 0.272~mag in $F160W$, $F555W$ and $F814W$. To accommodate the potential impact of the gyro drift in the photometry, we adopt a larger aperture of 5 pixels (instead of 3 pixels) for the images that are the most affected by the drift (indicated by (*) in Table \ref{table:observations}). We measured the ``instant" drift for each image by comparing the changes in position for the Cepheid between successive exposures and we found 13 Cepheids with a drift larger than 0.10\arcsec\ during an exposure in at least one filter (this threshold corresponds to a drift of 0.8 pixels in NIR and 2.5 pixels in UVIS images). For the 5 pixel aperture, we adopt aperture corrections of 0.154, 0.074 and 0.078~mag in $F160W$, $F555W$ and $F814W$ respectively \citepalias{Riess2019}.

In five UVIS images (``iev9a4hpq", ``iev9a4hqq", ``iev9c2v3q", ``iev9c2v4q", ``iev9tyd8q"), the Cepheid falls close to x=1024, where there is a detector defect. Vertical lines at x=1023, x=1024 and x=1025 are respectively fainter, brighter and brighter than the median of the background by a significant amount. For these images, we corrected for this effect by subtracting the median value of each line (this changes the Cepheid magnitude by 0.010~mag at most).

For UVIS binned images, the {\textit{HST}} pipeline does not produce CTE-corrected images (``FLC"); only the standard calibrated images (``FLT") are available. For these images, we applied a small term of 0.015~mag determined from unbinned images to account for the absence of CTE correction, however this does not affect the main results since it almost always applies to $F555W$ and $F814W$ simultaneously, and therefore cancels out in the $m_H^W$ Wesenheit index (Eq. \ref{eq:wesenheit}, see \S\ref{sec:PLR}). We note that this correction might slightly affect the optical Wesenheit magnitude (see Eq. \ref{eq:wesenheit_opt}) but the impact in the P--L relation is nearly negligible. 
Finally, following \citetalias{Riess2019}, we correct for the small difference in the WFC3-UVIS photometry between Shutters A and B: we add/subtract a small term for Shutter A/B respectively, where this term is 0.006 mag in $F555W$ and 0.0035 in $F814W$ \citep{Sahu2014}. \\

\begin{table*}[t!]
\caption{Mean magnitudes of our sample of SMC Cepheids. \\}
\footnotesize
\centering
\begin{tabular}{c c c c c c c c c c c c c}
\hline
\hline
Cepheid & R.A. & Dec. & Geo$^{\, (a)}$  & $\log P$ & $F555W$ & $\sigma$ & $F814W$ & $\sigma$ & $F160W^{\, (b)}$ & $\sigma$ & $m_H^{W\, (c)}$ & $\sigma$  \\
 & \multicolumn{2}{c}{\scriptsize{(deg)}}  & \scriptsize{(mag)} & \scriptsize{(d)}& \multicolumn{2}{c}{\scriptsize{(mag)}} & \multicolumn{2}{c}{\scriptsize{(mag)}}& \multicolumn{2}{c}{\scriptsize{(mag)}}& \multicolumn{2}{c}{\scriptsize{(mag)}}\\
\hline
OGLE-0518 & 10.801423 & -73.325446 & -0.089 & 1.198 & 15.385 & 0.017 & 14.224 & 0.008 & 13.161 & 0.027 & 12.653 & 0.028  \\
OGLE-0524 & 10.828177 & -73.338793 & -0.089 & 1.022 & 15.491 & 0.012 & 14.520 & 0.010 & 13.693 & 0.028 & 13.259 & 0.028  \\
OGLE-0541 & 10.873654 & -73.002488 & -0.052 & 1.137 & 15.149 & 0.018 & 14.077 & 0.011 & 13.088 & 0.027 & 12.651 & 0.028  \\
OGLE-0570 & 10.947143 & -73.240723 & -0.075 & 1.037 & 15.425 & 0.011 & 14.435 & 0.008 & 13.496 & 0.027 & 13.068 & 0.028  \\
OGLE-0576 & 10.963393 & -73.332920 & -0.084 & 1.159 & 15.244 & 0.011 & 14.132 & 0.008 & 13.107 & 0.027 & 12.623 & 0.028  \\
OGLE-0668 & 11.157914 & -73.136825 & -0.056 & 0.906 & 15.318 & 0.017 & 14.448 & 0.009 & 13.669 & 0.028 & 13.307 & 0.029  \\
OGLE-0672 & 11.165211 & -73.263262 & -0.069 & 0.948 & 15.627 & 0.010 & 14.634 & 0.009 & 13.673 & 0.028 & 13.250 & 0.028  \\
OGLE-0694 & 11.223210 & -73.289953 & -0.069 & 0.961 & 16.196 & 0.016 & 15.021 & 0.012 & 13.885 & 0.028 & 13.391 & 0.029  \\
OGLE-0705 & 11.264848 & -73.307750 & -0.070 & 1.021 & 15.624 & 0.017 & 14.737 & 0.012 & 13.699 & 0.028 & 13.316 & 0.029  \\
\hline
\end{tabular}
{\flushleft \textbf{Notes:} \\ (a) The geometric correction is derived using Eqns.~\ref{eq:geocorr_mag} and \ref{eq:geocorr_final}. \\
(b) Does not include CRNL correction (0.0293 $\pm$ 0.0023~mag or 3.8~dex) between SMC and extragalactic Cepheids. \\ (c) Includes geometric correction and CRNL correction (0.0293 $\pm$ 0.0023~mag or 3.8~dex) between SMC and extragalactic Cepheids.\par}
{\center (This table is available in its entirety in machine-readable form.) \\ }
~ \\
~ \\
\label{table:mean_mags}
\end{table*}

\subsection{Phase corrections} 

Our random-phase single-epoch \textit{HST} measurements were transformed into intensity-averaged magnitudes using the well covered $V$ and $I$ light curves from the OGLE-IV survey \citep{Soszynski2015}, obtained with the 1.3-m Warsaw telescope at Las Campanas Observatory in Chile. These light curves were supplemented by additional data points obtained more recently with the same telescope and camera after the release of the OGLE-IV catalog and include the most recent photometric measurements through July 2023 (I. Soszy\'nski, private communication, 2023). A few outlier points were rejected from the $V$-band light curves for OGLE-1117, 2437 and 0705, and from the $I$-band light curves for OGLE-1291, 1477 and 1157. These light curves have an average of 59 and 1,118 data points per object in $V$ and $I$, respectively. In the case where the $V$-band light curve has less than 10 data points (OGLE-2488, 1025, and 1399), we derive it from the $I$-band light curve assuming the phase-lag $\phi_I = \phi_V + 0.023$ and the amplitude ratio $A_I = 0.6 \, A_V$, both derived from our sample of OGLE light curves. Except for these three Cepheids, the minimum number of points is 19 in $V$ and 208 in $I$.
We fit the OGLE light curves with Fourier series to create a model from which to estimate the phase corrections and we adapt the number of Fourier modes between 4 and 8 to minimize the $\chi^2$ of the fit. Finally, for the $F160W$ filter, we adopted the $H$-band templates and phase lag from \cite{Inno2015}.

The periods of our Cepheids were taken from OGLE-IV and the phases of our \textit{HST} observations are derived from these periods. The phase corrections were derived using the same approach as described in \cite{Riess2018a} and are shown in Fig.~\ref{fig:phase_corrections}. We find a mean difference of -0.002, -0.005 and 0.010~mag and standard deviations of 0.09, 0.25 and 0.15~mag in $F160W$, $F555W$ and $F814W$ respectively. For comparison, \citetalias{Riess2019} found differences of -0.001, -0.048 and -0.013~mag with standard deviations of 0.11, 0.29 and 0.17~mag in $F160W$, $F555W$ and $F814W$ respectively.  \\

\section{Period-Luminosity relations} 
\label{sec:PLR}

\subsection{SMC P--L relations} 

The final intensity averaged magnitudes are listed in Table \ref{table:mean_mags}. We combine the three filters to construct the Wesenheit index $m_H^W$ defined as:
\begin{equation}
m_H^W = F160W - 0.386 \, (F555W - F814W)
\label{eq:wesenheit}
\end{equation}
where the color coefficient 0.386 is derived from the \cite{Fitzpatrick1999} reddening law assuming $R_V=3.3$, following \cite{Riess2022a}. Another secondary quantity used in this work is the optical Wesenheit index, derived under the same assumptions as:
\begin{equation}
m_I^W = F814W - 1.19 \, (F555W - F814W)
\label{eq:wesenheit_opt}
\end{equation}

\begin{figure*}[t!]
\centering
\includegraphics[width=17.8cm]{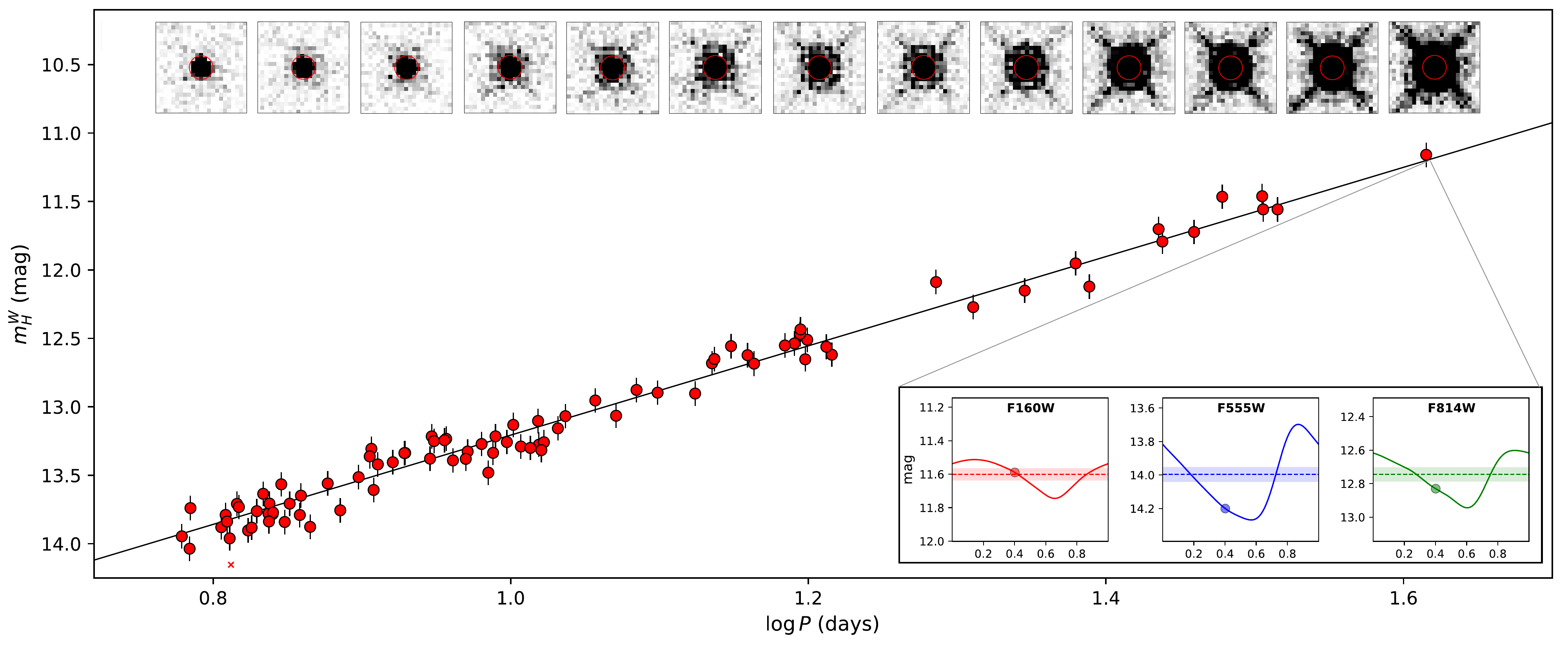}
\caption{P--L relation in the $m_H^W$ Wesenheit index for our sample of SMC Cepheids. Postage stamps in $F160W$ are displayed for a number of Cepheids with increasing periods: the stamps are on the same scale with a size of 1.5\arcsec\ , and the red circles are the 3 pixel apertures. An example of light curve fit is shown for the longest-period Cepheid (OGLE-1797) in the three filters. The outlier (OGLE-1455 at $3.4\sigma$) is shown with a red cross symbol. \\}  
\label{fig:Large_PL_with_stamps}
\vspace{0.7cm}
\end{figure*}

We correct for the count rate non-linearity (CRNL) of the WFC3/IR detector, seen as a dimming of fainter sources, which results from photons at low count rates being detected less efficiently than photons at high count rate \citep[0.0077 $\pm$ 0.0006~mag/dex,][]{Riess2019CRNL}: we add 0.0293 $\pm$ 0.0023~mag to $F160W$ measurements, corresponding to a 3.8~dex flux ratio\footnote{\citetalias{Riess2019} assumed a 4 dex flux ratio between LMC Cepheids ($H \sim 13$ mag) and the sky-dominated Cepheids observed in SNe~Ia hosts and NGC 4258 ($H \sim 23$ mag): the flux is defined as $F=10^{(25-m)/2.5}$ so $F_{LMC} = 63,000$ and $F_{N4258} = 6.3$. The flux ratio is therefore $10^4$, hence 4 dex. For SMC Cepheids ($H \sim 13.5$ mag) we follow the same procedure and obtain $F_{SMC} = 39,810$, therefore a flux ratio of $10^{3.8}$, hence 3.8 dex.} between SMC Cepheids and Cepheids observed in SN~Ia host galaxies and NGC$\,$4258. By convention, this is applied to anchor Cepheids rather than those in SN~Ia hosts. We also add a small corrective term to account for depth effects in the SMC (see 4th column in Table \ref{table:mean_mags}). This geometric correction is discussed in details in \S\ref{sec:geocorr}. Finally, we follow \citetalias{Riess2019} and account for the width of the instability strip by adding the following quantities in quadrature to the magnitude uncertainties: 0.07~mag, 0.08~mag, 0.09~mag, 0.14~mag and 0.22~mag in $m_H^W$, $m_I^W$, $F160W$, $F814W$, and $F555W$ respectively.

We adopt a P--L slope of $-3.26 \, \rm mag/dex$ in $m_H^W$ from \citetalias{Riess2019} for consistency, and fit the P--L intercept with a Monte Carlo algorithm. Following \citetalias{Riess2019} we perform a sigma-clipping with a threshold of $2.75 \sigma$ from Chauvenet’s criterion. After rejecting one outlier (OGLE-1455, found at $3.4 \sigma$), we obtain a P--L scatter of $0.1017 \, \rm mag$ (after including the geometry correction described in \S\ref{sec:geocorr} and Eq.~\ref{eq:geocorr_final}). The dispersion is slightly higher than in the LMC which we attribute to an $\sim$0.05 mag dispersion along the line of sight,  uncorrected by our simple geometric model, and we combine that in quadrature to the errors for the final errors. The P--L relation is shown in Fig.~\ref{fig:Large_PL_with_stamps}, together with postage-stamp cutout images of Cepheids in $F160W$, as well as an example of light curve fit. Thanks to the higher resolution of HST and to the use of a Cepheid sample in the SMC core, we avoid crowding and depth effects and obtain a P--L dispersion that is significantly lower than the ground-based P--L relation derived by the OGLE collaboration in the $W_{VI}$ index \citep[0.155~mag,][thus far the tightest P--L relation in the SMC, ]{Soszynski2015}.

\begin{figure*}[t!]
\centering
\includegraphics[width=8.5cm]{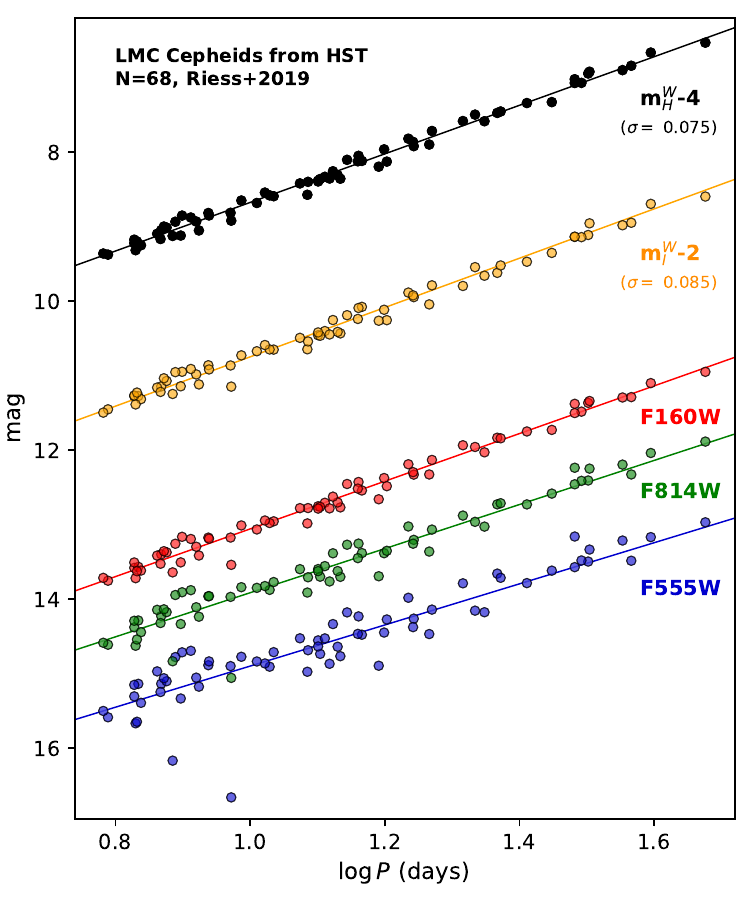} ~~~
\includegraphics[width=8.5cm]{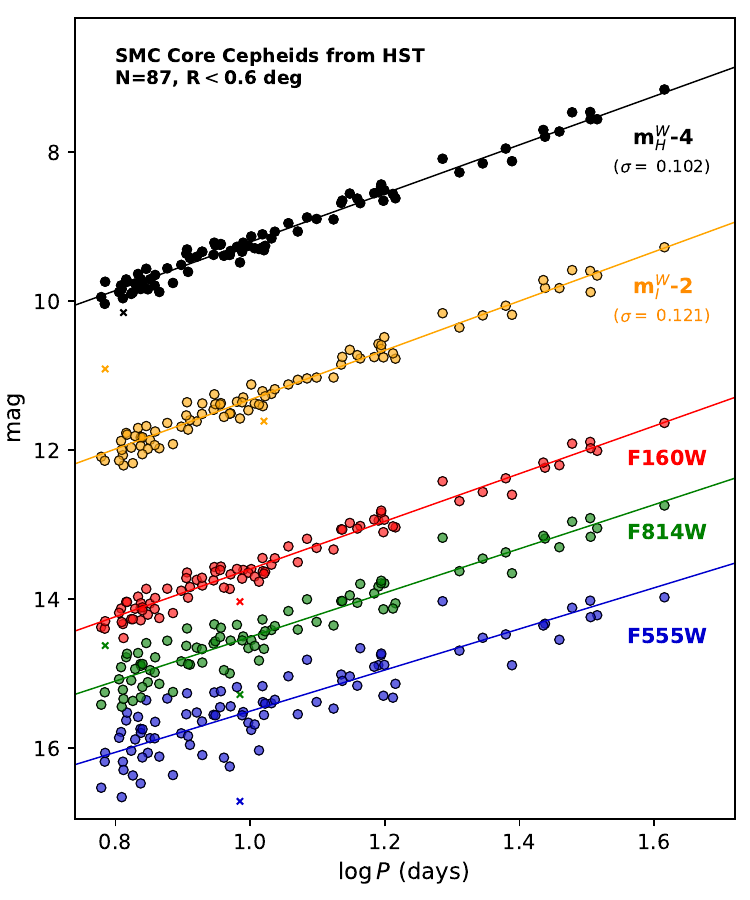}
\caption{Period-Luminosity relations in the 3 \textit{HST} filters and in the $m_I^W$ and $m_H^W$ Wesenheit indices. Outliers are shown by small ``x" symbols and are excluded from the fit. Left and right panels show the Period-Luminosity relations in the LMC \citep{Riess2019} and SMC respectively. \\ 
\label{fig:All_PL}}
\end{figure*}

The P--L relations obtained in $F160W$, $F555W$, $F814W$, $m_H^W$ and $m_I^W$ are shown in Fig.~\ref{fig:All_PL}, where the slopes are fixed to the \citetalias{Riess2019} values. Table \ref{table:PL_results} gives the free slopes derived from our sample and the P--L intercepts obtained with the fixed slopes from \citetalias{Riess2019}. The $m_H^W$ free slope is $-3.31 \pm 0.05 \, \rm mag/dex$, consistent to within $1 \sigma$ with the $-3.26$ slope from \citetalias{Riess2019}. In the other filters, our slopes are also in excellent agreement with the slopes from \citetalias{Riess2019}. Recent studies have suggested a possible break in the P--L relation \citep{Bhardwaj2016b, Kodric2018, Bhardwaj2020}: with this low-dispersion data we find no evidence for a break in the P--L relation in the LMC and SMC as presented in Appendix \ref{apd_B}.

\begin{table}[t!]
\caption{Cepheid P--L relations from \textit{HST} photometry in the SMC: $m_H^W = \alpha \log P + \beta$.  }
\footnotesize
\centering
\begin{tabular}{l c c c c c c}
\hline
\hline
Filter & $\alpha$ & $\alpha$ & $\beta$ & $\sigma$   \\
     & (this work) & \citepalias{Riess2019} & \\
\hline
$F555W$ & $-2.65 \pm 0.13$ & $-2.76 \pm 0.13$ & $18.263 \pm 0.031$ & 0.287 \\
$F814W$ & $-2.98 \pm 0.09$ & $-2.96 \pm 0.09$ & $17.467 \pm 0.021$ & 0.195 \\
$F160W$ & $-3.22 \pm 0.06$ & $-3.20 \pm 0.04$ & $16.795 \pm 0.013$ & 0.122 \\
$m_I^W$ & $-3.37 \pm 0.06$ & $-3.31 \pm 0.04$ & $16.632 \pm 0.013$ & 0.121 \\
$m_H^W$ & $-3.31 \pm 0.05$ & $-3.26 \pm 0.04$ & $16.467 \pm 0.011$ & 0.102 \\
\hline
\end{tabular}
~ \\
{\flushleft \textbf{Notes:} Intercepts in column 4 are derived with slopes fixed to \citetalias{Riess2019} values obtained in the LMC (column 3). All intercepts include geometric corrections and intercepts in $F160W$ and $m_H^W$ include CRNL. \\ \par}
~ \\
\label{table:PL_results}
\end{table}

The $m_H^W$ Wesenheit index defined in Eq.~\ref{eq:wesenheit} is constructed assuming the \cite{Fitzpatrick1999} reddening law with $R_V=3.3$, which gives a coefficient of $R=0.386$. However \cite{Gordon2003} suggested that the SMC dust corresponds better to $R_V = 2.74$, which would give $R = 0.362$. Using this alternate definition of $ m_H^W $ yields a slightly larger P--L scatter of $0.1024 \, \rm mag$. Using $R_V=3.3$ ensures consistency with the SH0ES distance ladder based on the same assumption. As explained in Appendix D of \cite{Riess2022a}, it is not valid to use a different reddening law in different hosts by varying $R$: this would require to first derive a period-color relation for SMC Cepheids and to remove the intrinsic color from the apparent color. While $R_V = 3.3$ corresponds to a reddening term at $1.6 \, \mu \rm m$ of 0.386 for the \cite{Fitzpatrick1999} law, $R_V = 2.7$ corresponds to 0.362, a difference of only 0.02. After removing an intrinsic, period-color relation, the color excess, $E(V-I)$ for the SMC is $\sim0.1 \, \rm mag$ so that a change in the reddening law would shift our results by only $\sim 0.002$ mag. \\

\newpage
\subsection{The geometry of the SMC} 
\label{sec:geocorr}

The SMC is known to have an elongated structure, which can complicate distance measurements based on this dwarf galaxy. From a sample of Cepheids, \cite{Mathewson1988} found a significant line-of-sight depth of about 20~kpc (compared to its distance of $\sim 62 \, \rm kpc$) and that the northeastern section of the bar is $10-15 \, \rm kpc$ closer than the southern section. \cite{Nidever2013} also showed that the optical depth in the eastern part of the SMC is two times higher than in the western part, and that the eastern part comprises of two groups of stars with different mean distances. More recently, \cite{Murray2023} presented evidence that the SMC is actually composed of two substructures with distinct chemical compositions separated by $\sim 5 \, \rm kpc$ along the line of sight. As a result, any study of SMC Cepheids must include their non-negligible spread in distance. 

In order to mitigate the impact of its elongated shape, a first solution is to limit the sample to a narrow region in the SMC core. \cite{Breuval2022} showed that the ideal region (based on reducing the P--L scatter) is a radius of $0.6 \, \rm deg$ around the SMC center, where the number of Cepheids is still sufficient to populate the P--L relation, while the residual geometry effects are minimized. For example, adopting a Cepheid sample in a region of $R\!=\!0.6 \, \rm deg$ compared to a larger region of $R\!=\!2.5 \, \rm deg$ reduces the P--L scatter by $0.02 \, \rm mag$ (see Fig.~\ref{fig:PL_scatter_vs_radius}) and changes the P--L intercept by $0.050 \, \rm mag$, so that extending to the larger radius results in a significant degradation in accuracy with respect to the uncertainties reported in Table~\ref{table:PL_results}.

Additionally, one can attempt to correct for the differential depths of each Cepheid by adopting a planar model for the SMC. For example, \cite{Breuval2021, Breuval2022} corrected for the SMC depth by adopting the planar model described by DEBs from \cite{Graczyk2020}. The corrected distance to each Cepheid is: 
\begin{equation}
d(x, y) =  +3.086 \, x -4.248 \, y  + d_{\rm SMC} 
\label{eq:geocorr_G20}
\end{equation}
where $(x, y)$ are the cartesian coordinates of the Cepheids obtained from their equatorial coordinates $(\alpha, \delta)$ \citep[see][\S3.3 for the transformations]{Breuval2021}, considering the SMC center at ($12^{\circ}.54$, $-73^{\circ}.11$) from \cite{Ripepi2017}, and $d_{\rm SMC}\!=\!62.44 \pm 0.47 \pm 0.81 \, \rm kpc$ is the mean SMC distance from \cite{Graczyk2020}, based on DEBs. The geometric correction we apply to each Cepheid, $\rm mag_{corr}$, is derived with:
\begin{equation}
\rm mag_{\rm corr} = 5 \, \log_{10} (d_{SMC} / d(x, y))
\label{eq:geocorr_mag}
\end{equation}
This correction is valid as long as Cepheids and DEBs share the same center and have a similar spatial distribution, which is true near the SMC center but may not be the case in the outskirts. Also, the SMC geometry is not planar at large scales, but this is a good approximation for the core region. Using this correction we obtain a P--L scatter of 0.1022~mag. We then refined this geometry model based on our own sample of Cepheids: we varied both coefficients and retained the solution that minimizes the $\chi^2$ of the P--L relation. We find an optimal geometry of:
\begin{equation}
d(x, y) = +3.480 \, x -2.955 \, y  + d_{\rm SMC} 
\label{eq:geocorr_final}
\end{equation}
which gives a final P--L scatter of 0.1017~mag. We note that the uncertainties of the geometric coefficients are of the order of $\pm 1$. The difference between the coefficients in Eq. \ref{eq:geocorr_G20} and \ref{eq:geocorr_final} can be attributed to the use of different tracers (15 DEBs vs 88 Cepheids) and to their different distributions (extended SMC vs core region). These corrections have a mean value of $-0.003 \, \rm mag$ and a dispersion of $0.047 \, \rm mag$, with a minimum and maximum of $-0.088 \, \rm mag$ and $+0.098 \, \rm mag$. 
We can verify the agreement of our geometric model with the literature by overplotting it with Fig. 4 from \cite{Graczyk2020}, based on the distribution of 15 SMC DEBs, with Fig. 6 from \cite{Scowcroft2016}, obtained from \textit{Spitzer} mid-infrared observations of 90 Cepheids, with Fig. 17 from \cite{Ripepi2017}, from the NIR light curves of 4793 SMC Cepheids observed with the VMC survey, and with Fig. 15 from \cite{Jacyszyn2016} derived from optical light curves of 2646 fundamental-mode Cepheids from the OGLE survey. These comparisons show a good agreement between Eq.~\ref{eq:geocorr_final} and the SMC core-region geometry in these studies.

In this section, we showed how the considerable line-of-sight depth of the SMC might affect the P--L relation and how we can greatly mitigate its impact for distance measurements. The SMC geometry is still challenging and its structure must be better understood \citep{Murray2023}, but for our purposes, the present Cepheid-based geometric corrections and core-region sample best reduce their impact, and any residual effects only slightly increase the P--L scatter.  \\

\begin{figure}[t!]
\includegraphics[width=8.2cm]{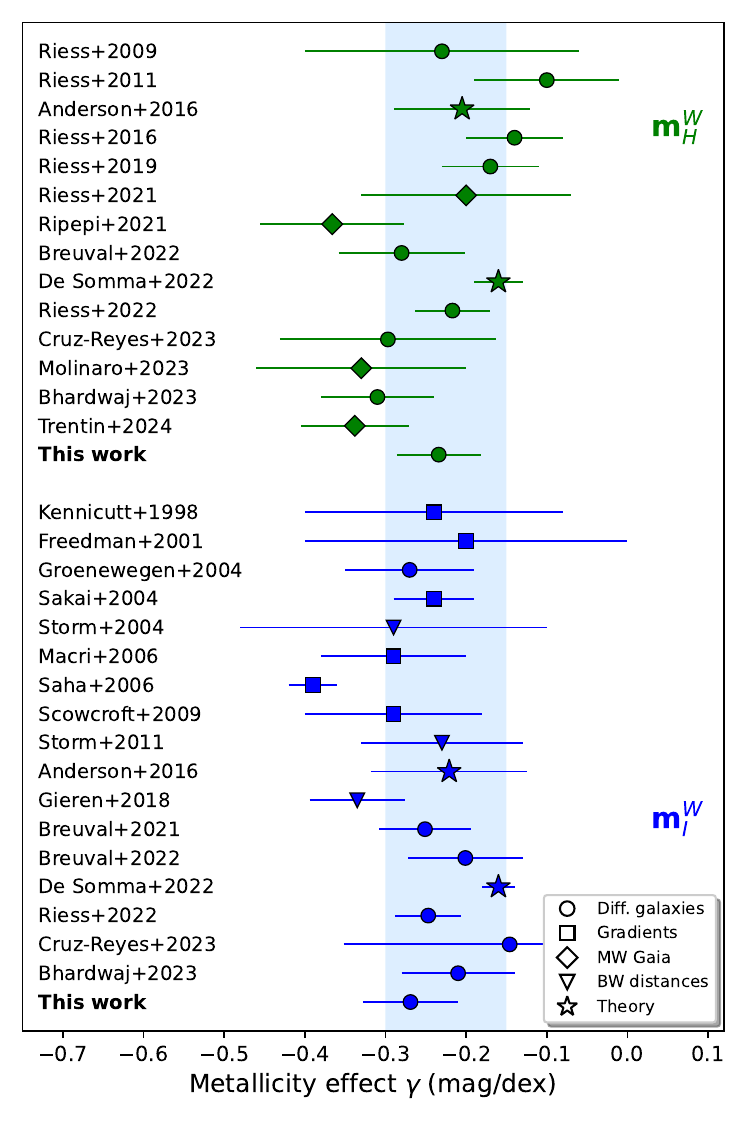}
\caption{Metallicity dependence in $m_H^W$ and $m_I^W$ from the literature. Some issues were identified in the results by \cite{Wielgorski2017}, \cite{Freedman2011} and \cite{Madore2024}, therefore they are not shown in this figure \citep[see discussion in][section 5.5]{Breuval2022}. The blue region covers $-0.15$ to $-0.30$ mag/dex, where most values are found.
\label{fig:gamma_literature}}
\end{figure}

\subsection{Calibration of the metallicity effect in the \textit{HST}/WFC3 photometric system} 
\label{sec:gamma}

The Cepheid metallicity effect has been a source of intense study for over three decades \citep{Freedman1990, Freedman2001, Sakai2004}. In previous-generation measurements of the Hubble constant, which relied on comparisons between Cepheids in the low-metallicity LMC and metal-rich spirals, this term played a significant role. More recent determinations of H$_0$ make use of metal-rich calibrations in NGC$\,$4258 and from improved Milky Way parallaxes, thus reducing the full impact of this term to $<$ 1 km/s/Mpc.  

In the regime covered by SNe~Ia host galaxies (with metallicities between Milky Way and LMC Cepheids), there is a consensus for its sign and value ($\gamma\!\sim\!-0.25 \, \rm mag/dex$) thanks to improved distances and metallicities \citep[][see Figure \ref{fig:gamma_literature} for a literature summary]{Breuval2022, Romaniello2022, Trentin2024}. However, this term is still relatively weakly constrained in the metal-poor regime ($\rm [Fe/H] < -0.5 \, dex$). 
Its calibration over a larger metallicity range requires combining precise abundance measurements in multiple galaxies with homogeneous Cepheid photometry in a consistent system. Our sample of SMC Cepheids with HST/WFC3 photometry represents a unique opportunity to extend the calibration of this effect over a metallicity range twice as large as that currently available.

We compile three samples all with identical-system photometry: 
\begin{itemize}
\item 67 Milky Way Cepheids from \cite{Riess2021},
\item 70 LMC Cepheids from \citetalias{Riess2019},
\item 87 SMC Cepheids from the present paper. 
\end{itemize}
The size of the three samples is defined by the number of Cepheids with HST/WFC3 photometry after excluding outliers. The Cepheids of the Milky Way in our sample are all located in the vicinity of the Sun. Their metallicities are taken from Bhardwaj et al. (2023 ) and were measured recently using high-resolution spectra obtained with ESPaDOnS on CFHT. The mean metallicity of our Milky Way sample is $+0.033 \pm 0.013 \, \rm dex$ and the dispersion is $0.11 \, \rm dex$.

\begin{figure}[t!]
\includegraphics[width=8.4cm]{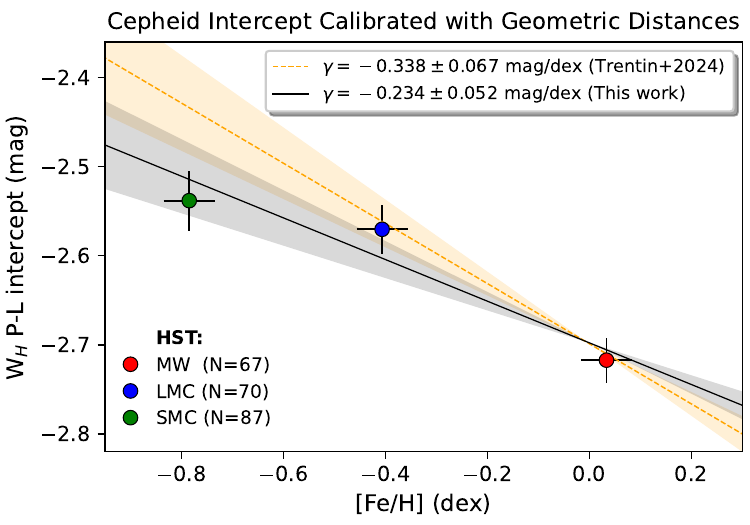}
\caption{Wesenheit $m_H^W$ P--L intercept in absolute magnitudes (at $\log P =0$) in the Milky Way, LMC and SMC based on HST photometry (red, blue and green circles respectively). The orange and black lines represent the metallicity effect from \citet{Trentin2024} (their Table 4, line 41) and from the present work. The shaded regions show the uncertainties in $\gamma$ [Fe/H].}
\label{fig:intercept_metallicity}
\end{figure}

For LMC Cepheids, we adopt the metallicities from \cite{Romaniello2022} obtained with UVES on the ESO VLT, with individual measurements for all Cepheids and a mean value of $\rm [Fe/H] = -0.409 \pm 0.003 \, dex$, and a dispersion of $0.076 \, \rm dex$. Finally, we adopt the new metallicity measurements of SMC Cepheids from Romaniello et al.~(2024, in prep.) which include 44 Cepheids of our sample, an average value of $\rm [Fe/H]\!=\!-0.785 \pm 0.012 \, dex$, with a dispersion of $0.082 \, \rm dex$, and we adopt this average value also for the other SMC Cepheids not included in Romaniello et al. sample. 

To measure the metallicity effect, we first derive the absolute magnitude for each Cepheid of the three samples assuming their geometric distances (\textit{Gaia} DR3 parallaxes of Milky Way Cepheids, including the \cite{Lindegren2021_plx_bias} zero-point and the residual $14\, \rm \mu as$ correction determined from Cepheids by \citet{Riess2021}, and the LMC and SMC distances from eclipsing binaries, respectively $49.59 \pm 0.09 \pm 0.54$~kpc and $62.44 \pm 0.47 \pm 0.81$~kpc). We fit the absolute P--L relation in each galaxy with a fixed slope of $-3.26 \, \rm mag/dex$. Then, we fit a linear relation between the intercept and the mean metallicity \citep[see method in][]{Breuval2022}: $\beta = \gamma \, \rm [Fe/H] + \delta$ (solid black line in Fig.~\ref{fig:intercept_metallicity}). In the three galaxies, we conservatively assume an uncertainty of $0.05 \, \rm dex$ for the mean metallicity and we add the distance uncertainties in quadrature to the intercept errors. We obtain $\gamma\!=\!-0.234 \pm 0.052 \, \rm mag/dex$, where the uncertainties are estimated via a Monte Carlo algorithm. In the optical ($m_I^W$), we find $\gamma = -0.264 \pm 0.058 \, \rm mag/dex$.

These values are consistent with other recent measurements based on a combination of geometric distances in the Milky Way and Magellanic Cloud Cepheids \citep{Breuval2021, Breuval2022, Bhardwaj2023, CruzReyes2023}. Independent calibrations of the metallicity effect based exclusively on several hundred Milky Way Cepheids with \textit{Gaia} DR3 parallaxes from the C-MetaLL project\footnote{\href{https://sites.google.com/inaf.it/c-metall/home}{https://sites.google.com/inaf.it/c-metall/home}} including new spectra of very metal poor Cepheids have resulted in stronger effects of the order of $-0.4 \, \rm mag/dex$ in the \textit{Gaia} $W_G$, optical $W_{VI}$ and NIR $W_{VK}$, $W_{JK}$ Wesenheit indices \citep{Ripepi2022, Trentin2024, Bhardwaj2024}. However, when applying the parallax zero-point counter-correction of $14 \, \rm \mu as$ from \cite{Riess2021}, they recover the canonical result of $\sim -0.25$ to $-0.35$~mag/dex in the specific $W_H$ Wesenheit index used in the SH0ES distance ladder and good agreement with the LMC DEB distance \cite[see][Figure 16]{Trentin2024}.

\begin{table*}[t!]
\caption{Uncertainty in H$_0$ from leading term, geometric calibration of Cepheids (\%).}
\centering
\begin{tabular}{l l c c c c c c c}
\hline
\hline
\multicolumn{1}{c}{Term} & \multicolumn{1}{c}{Description} & \multicolumn{3}{c}{\cite{Riess2022a, Riess2022b}} & \multicolumn{4}{c}{This paper} \\
     &             & LMC & MW & N4258                        & LMC & MW & N4258 & SMC \\
\hline
$\sigma_{\rm \mu, \, anchor}$  & Anchor distance                   & 1.2 & 0.8$\, ^{(a)}$ & 1.5$\, ^{(b)}$ & 1.2 & 0.8$\, ^{(a)}$ & 1.5$\, ^{(b)}$ & 1.5 \\
$\sigma_{\rm PL, \, anchor}$   & Mean of P--L in anchor             & 0.4 & ...            & 1.0            & 0.4 & ...            & 1.0            & 0.5 \\
$R \sigma_{\rm \lambda, 1, 2}$ & Zeropoints, anchor-to-host        & 0.1 & 0.1$\, ^{(a)}$ & 0.0            & 0.1 & 0.1$\, ^{(a)}$ & 0.0            & 0.1 \\
$\sigma_{\rm Z}$               & Cepheid metallicity, anchor-hosts & 0.4 & 0.15           & 0.15           & 0.4 & 0.1            & 0.1            & 0.5 \\
\hline
                               & Subtotal per anchor               & 1.4 & 0.8            & 1.8            & 1.4 & 0.8            & 1.8            & 1.7 \\
                               &                                   & \multicolumn{3}{p{2.9cm}}{\raisebox{.66\baselineskip}{$\underbrace{\hspace{3.0cm}}$}} & \multicolumn{4}{p{4.0cm}}{\raisebox{.66\baselineskip}{$\underbrace{\hspace{4.2cm}}$}}     \\
First Rung Total                               &                                   & \multicolumn{3}{c}{0.65}              & \multicolumn{4}{c}{0.60}  \\
\hline
\end{tabular}
 ~ \\
\tablecomments{\, (a) Includes both field Cepheids and cluster Cepheids; (b) \cite{Reid2019}. }
\vspace{0.3cm}
\label{table:error_budget}
\end{table*}

An outlier to this consensus comes from one study by \cite{Madore2024}, which selects Cepheid calibrations from tip of the red giant branch (TRGB) distances rather than the geometric anchors, comparing 28 hosts and concluding that $\gamma \sim 0$ for $m_I^W$. However, measuring the Cepheid metallicity effect by comparing Cepheid and TRGB distances requires that TRGB distances be independent of metallicity, which is not certain \citep[see for example][]{Wu2022}. More importantly, the Cepheid data for this study rely heavily on older and relatively inhomogeneous data sources (21 different photometric systems) and smaller samples which have been superceded in recent studies. Compared to the combination of {\it HST} measurements of Cepheids in the SMC, LMC and Milky Way (which produced $\gamma = -0.264 \pm 0.058 \, \rm mag/dex$ for $m_I^W$ here), \cite{Madore2024} exclude all Milky Way Cepheids (and \textit{Gaia} parallaxes) and find at face value $\gamma = -0.096 \, \rm mag/dex$ between the SMC and LMC, only half the size of our result. However, their SMC sample is based on only 9 fundamental-mode Cepheids, while the OGLE survey provides more than 2,300 Cepheids in the SMC \citep[][or the 88 here identically calibrated to those in the LMC and MW]{Soszynski2015}.  Some of the evidence for their low absolute value of $ \gamma $ is based on the four hosts with the lowest metallicity, namely Sextans A and B, WLM and IC 1613; for these, the Cepheid metallicity estimate of $ \sim -1.3 $~dex is based on the host metallicity \citep{Sakai2004}. 
Spectra of young massive stars in WLM \citep[-0.5 to -1.0~dex with a mean of -0.87~dex,][]{Urbaneja2008} are more metal rich than that value of -1.3 dex used for that host.  If the metallicities of new massive stars like Cepheids in these metal poor hosts are greater than the host mean, this would lead to an underestimate of the metallicity term.  Obtaining spectra of Cepheids in these hosts would be warranted to resolve the issue. Our calibration of the metallicity effect, based on direct measurements for a large number of Cepheids, is in good agreement with previous distance-ladder studies by the SH0ES team and other independent works (see Fig.~\ref{fig:gamma_literature}).  \\


\section{The Hubble constant}
\label{sec:H0}


 \begin{table*}[t!]
\caption{SMC Period-Luminosity relations from the literature.}
\footnotesize
\centering
\begin{tabular}{c c c c c c c c c c}
\hline
\hline
Filter & Def. & Reference & Slope & Intercept & $\sigma$ & $N$ & $R_{\rm max}$ & Telescope &   \\
       &      &          & \scriptsize{(mag/dex)} & \scriptsize{(mag)} & \scriptsize{(mag)} &  & \scriptsize{(deg)}  \\
\hline
$V$ & & \cite{Soszynski2015}          & $-2.898_{\pm 0.018}$ & $17.984_{\pm 0.008}$ & 0.266 & 4747 & 3 (98\%) & Warsaw telescope,   \\
$I$  & & \cite{Soszynski2015}         & $-3.115_{\pm 0.015}$ & $17.401_{\pm 0.007}$ & 0.215 & 4940 & 6 (100\%) & Las Campanas & ground  \\
$m_{\rm I, \, grd}^W$ & (a) & \cite{Soszynski2015} & $-3.460_{\pm 0.011}$ & $16.493_{\pm 0.005}$ & 0.155 & 4743 &  & Observatory  \\
\hline
$G$   &     & \cite{Ripepi2023} & $-2.830_{\pm 0.033}$ & $17.757_{\pm 0.026}$ & 0.251 & 843 & 3 (99\%) & {\it Gaia} & space  \\
$m_G^W$ & (b) & \cite{Ripepi2023} & $-3.382_{\pm 0.021}$ & $16.592_{\pm 0.017}$ & 0.156 & 839 & 5 (100\%) &   \\
\hline
$J$ & & \cite{Ripepi2017}           & $-3.070_{\pm 0.026}$ & $16.778_{\pm 0.021}$ & 0.196 & 821 &      \\
$K_S$ & & \cite{Ripepi2017}         & $-3.224_{\pm 0.023}$ & $16.530_{\pm 0.018}$ & 0.167 & 821 & 3 (98\%)  & ESO/VISTA & ground   \\
$m_{JK}^W$ & (c) & \cite{Ripepi2017}  & $-3.334_{\pm 0.021}$ & $16.363_{\pm 0.017}$ & 0.158 & 821 & 6.5 (100\%) & telescope  \\
$m_{VK}^W$ & (d) & \cite{Ripepi2017}  & $-3.291_{\pm 0.021}$ & $16.375_{\pm 0.017}$ & 0.155 & 821 &       \\
\hline
$[3.6 \, \rm \mu m]$ & & \cite{Scowcroft2016} & $-3.306_{\pm 0.050}$ & $16.492_{\pm 0.017}$ & 0.161 & 90 & 3 (98\%) & {\it Spitzer} & space  \\
$[4.5 \, \rm \mu m]$ & & \cite{Scowcroft2016} & $-3.207_{\pm 0.060}$ & $16.357_{\pm 0.018}$ & 0.164 & 90 & 6.5 (100\%) &  \\
\hline
$m_I^W$ & (e) & {\bf This work} & $-3.31_{\pm 0.04}$ & $16.632_{\pm 0.013}$ & {\bf 0.121} & 87 & 0.6 & {\it Hubble} & space    \\  
$m_H^W$    & (f) & {\bf This work} & $-3.26_{\pm 0.04}$ & $16.467_{\pm 0.011}$ & {\bf 0.102} & 87 & 0.6 &     \\  
\hline
\end{tabular}
~ \\
{\flushleft 
\textbf{(a)}: $m_{\rm I, \, grd}^W = I -1.55 \, (V-I), \,$ \textbf{(b)}: $m_{G}^W = G -1.90 \, (BP-RP), \,$ \textbf{(c)}: $m_{JK}^W = K -0.69 \, (J-K), \,$ \textbf{(d)}: $m_{VK}^W = K -0.13 \, (V-K), \,$ \\ \textbf{(e)}: $m_{I}^W = F814W -1.19 \, (F555W-F814W), \,$  \textbf{(f)}: $m_H^W = F160W -0.386 \, (F555W-F814W)$. \\
\par}
\vspace{0.4cm}
\label{table:PL_SMC_literature}
\end{table*}

Since \cite{Riess2022a}, there have been several updates that could lead to an improved local measurement of H$_0 $. These include the cluster Cepheids from \cite{Riess2022b}, the pairwise SNe~Ia covariance analysis based on spectral similarities from \cite{Murakami2023}, and the revised metallicity measurements from \cite{Bhardwaj2023}. With the results presented here, we are now able to include the SMC as a bona fide fourth anchor of the distance ladder, in addition to (1) \textit{Gaia} DR3 parallaxes of Milky Way Cepheids and Cluster Cepheids \citep{Riess2021, Riess2022b}, (2) late-type DEBs in the LMC \citep{Pietrzynski2019, Riess2019}, and (3) the water-maser host galaxy NGC$\,$4258 \citep{Reid2019}. The SMC satisfies the two criteria established in \cite{Riess2022a} for the other 3 anchors: a geometric distance determination and Cepheid photometry measured on a homogeneous {\it HST} photometric system. We combine the 88 core SMC Cepheids measured here, the DEB distance from \cite{Graczyk2020}, and SMC Cepheid metallicity measurements from Romaniello et al.~(2024, in prep.) using the same formalism presented in \cite{Riess2022a}. We find H$_0\!=\!73.17 \pm 0.86$ \kms, a 1.2\% determination of the Hubble constant including systematic uncertainties. As shown in Table \ref{table:error_budget}, including the SMC reduces the error in the first rung from 0.65\% to 0.60\%. This reduction is slightly less than what would occur if all anchors had equal weight; however, the combined constraints from \textit{Gaia} MW field and cluster-hosted Cepheids have nearly double the precision of the SMC. Somewhat greater leverage comes from the inclusion of the SMC in the global determination of the metallicity term due to its low metallicity. With the SMC, the global fit of the distance ladder yields a metallicity dependence of $\gamma \, (m_H^W) = -0.190 \pm 0.047$~mag/dex.  This represents a 30\% improvement over the uncertainty of $\pm$0.067 mag/dex obtained without the SMC\footnote{We note for completeness that the lower metallicity uncertainty was already included in R22 thanks to the combined use of SMC and LMC ground-based data on the same photometric system, but without the value of the Cepheid geometric calibration.}. Unlike in \S\ref{sec:gamma}, here we use a covariance matrix to take into account that the LMC and SMC geometric distances have a common systematic error (their relative distance is measured better than their individual distances), which helps constrain the difference in Cepheid brightness. We note that the global metallicity scale used in the SH0ES distance ladder is based on [O/H], rather than [Fe/H] as used in the previous section.  This is most relevant when including low-metallicity hosts like the SMC, with their enhanced $\alpha/$Fe abundances; [O/Fe] is $\sim$ 0.2 for the SMC Cepheids (Romaniello et al.~2024, in prep.), higher than the value of $\sim$ 0.06 seen at higher metallicities such as in the MW \citep[e.g.][]{LuckLambert2011}. This somewhat compresses the [O/H] scale versus [Fe/H], leading to a smaller (in absolute value) metallicity term than from the earlier sections. With the SMC as the only anchor for the distance ladder, we obtain H$_0\!=\!74.1 \pm 2.1$ \kms, which shows the good consistency of the SMC with the other three anchors. \\ 

\section{Discussion}
\label{sec:discussion}

We presented the first photometric measurements of SMC Cepheids with the {\it Hubble} Space Telescope Wide Field Camera 3, using the same photometric system as the rest of the SH0ES distance scale. Combining the high resolution of \textit{HST} with a sample of 88 Cepheids in the core of the SMC, we mitigate the impact of crowding and of the SMC line-of-sight depth, and obtain the lowest dispersion for the Cepheid P--L relation in this galaxy to date: $\sigma = 0.102$~mag. Some recent P--L relations from the literature obtained in different photometric systems and filters are listed in Table \ref{table:PL_SMC_literature}. Thanks to the low metallicity of SMC Cepheids, we calibrate the metallicity dependence of Cepheids on a range twice larger than previous studies and we find good agreement with the literature with $\gamma = -0.234 \pm 0.052$ mag/dex, based on direct [Fe/H] measurements of Cepheids in the Milky Way, LMC and SMC. The global fit of the SH0ES distance ladder based on [O/H] metallicities gives a similar value, $\gamma = -0.190 \pm 0.047$ mag/dex. Since the [Fe/H] scale is more relevant to nearby galaxies like the SMC, we adopt $-0.234 \pm 0.052$ mag/dex as our final value. Although it would not directly impact the $H_0$ value, this dependence needs to be better calibrated in the most metal-poor regime beyond [Fe/H] $< -0.7$~dex, for example in Local Group galaxies WLM, IC1613, Sextans A and B using spectra of Cepheids there, and where possible non-linearities could be investigated.

In this work, we provide a new calibrating galaxy for the first rung of the distance scale with the first \textit{HST} photometric measurements of SMC Cepheids, increasing the number of geometric anchors from three to four, which yields a 1.2\% measurement of the Hubble constant, H$_0\!=\!73.20 \pm 0.86$ \kms. The small number of geometric anchors is primarily due to limitations in the late-type DEBs method. Eclipsing binaries are rare systems in which two stars orbit each other, and their orbital plane is aligned in such a way that each star regularly passes in front of the other, causing an eclipse. They present regular light curves with well-defined eclipses. ``Detached'' systems are well-separated and do not transfer mass between the components. The distance to these systems can be derived by comparing the components' linear diameter, obtained from the analysis of the photometric and radial velocity curves \citep{Guinan1998, Fitzpatrick2003}, and their angular diameter, measured using surface brightness-color relations. While late-type DEBs provide unprecedented precision for the geometric distances to the LMC and SMC, the downside of this method is that these systems are faint, typically $m_V \sim 18$~mag at the distance of the SMC, which makes them difficult to observe in more distant galaxies. On the other hand, early-type DEBs are much brighter and can reach $\sim$1 Mpc, including nearby galaxies M31 and M33 \citep{Bonanos2006}. The method to measure early-type DEB distances is different from the late-types as it relies on model atmosphere theory, instead of the 1\% precision surface brightness color relation used for late-type DEBs in the LMC and calibrated empirically \citep{Pietrzynski2019}. This relation is not yet calibrated on the color range applicable to early-type DEBs ($-1 < V-K < 0$). 
New results are expected from ongoing interferometric measurements of angular diameters for these stars in the solar neighborhood and in the LMC \citep{Challouf2014, Taormina2020}. This work is extremely important, as future geometric distances to M31 and M33 will directly provide two additional anchors for the distance scale, where Cepheid photometry is already available in the HST/WFC3 photometric system \citep{Li2021, Breuval2023}. 

Although the LMC geometric distance from \cite{Pietrzynski2019} remains the best DEB-based calibrator for the first rung of the distance ladder, the low metallicity and the large number of Cepheids in its core region make the SMC a powerful addition to the set. The SMC also has a historical importance as it is the galaxy where the P--L relation was first discovered and established by Henrietta Leavitt \citep{Leavitt1912}. Our consistent photometric measurements of Cepheids in the core of the SMC fall well within the distance ladder established from the three other anchor galaxies. With the addition of the SMC as a fourth anchor, the late universe measurement of the Hubble constant from the distance ladder moves closer to the goal of 1\% accuracy.


\newpage
\section*{Acknowledgements} 

We are grateful to the OGLE team for providing us with the most recent light curves for our sample of SMC Cepheids. We thank Erasmo Trentin and the C-MetaLL team for sharing their sample of Milky Way Cepheids. L.B. would like to thank Boris Trahin, Laura Herold, Siyang Li, Javier H. Minniti, and Richard I. Anderson for helpful discussions. Support for this work was provided by the National Aeronautics and Space Administration (NASA) through program No.~GO-17097 from the Space Telescope Science Institute (STScI), which is operated by AURA, Inc., under NASA contract No.~NAS 5-26555. This research is based primarily on observations with the NASA/ESA \textit{Hubble} Space Telescope, obtained at STScI, which is operated by AURA, Inc., under NASA contract No.~NAS 5-26555. This research was funded in part by the National Science Centre, Poland, grant no. 2022/45/B/ST9/00243. This research was supported by the Munich Institute for Astro-, Particle and BioPhysics (MIAPbP), which is funded by the Deutsche Forschungsgemeinschaft (DFG, German Research Foundation) under Germany's Excellence Strategy – EXC-2094 – 390783311. The data presented in this paper were obtained from the Mikulski Archive for Space Telescopes (MAST) at the Space Telescope Science Institute. The specific observations analyzed can be accessed via \href{https://archive.stsci.edu/doi/resolve/resolve.html?doi=10.17909/08t1-3x45}{doi:10.17909/08t1-3x45}. \\

\newpage
\appendix

\section{A break in the Cepheid P--L relation?}
\label{apd_B}

\begin{figure}[h!]
\centering
\includegraphics[width=12.5cm]{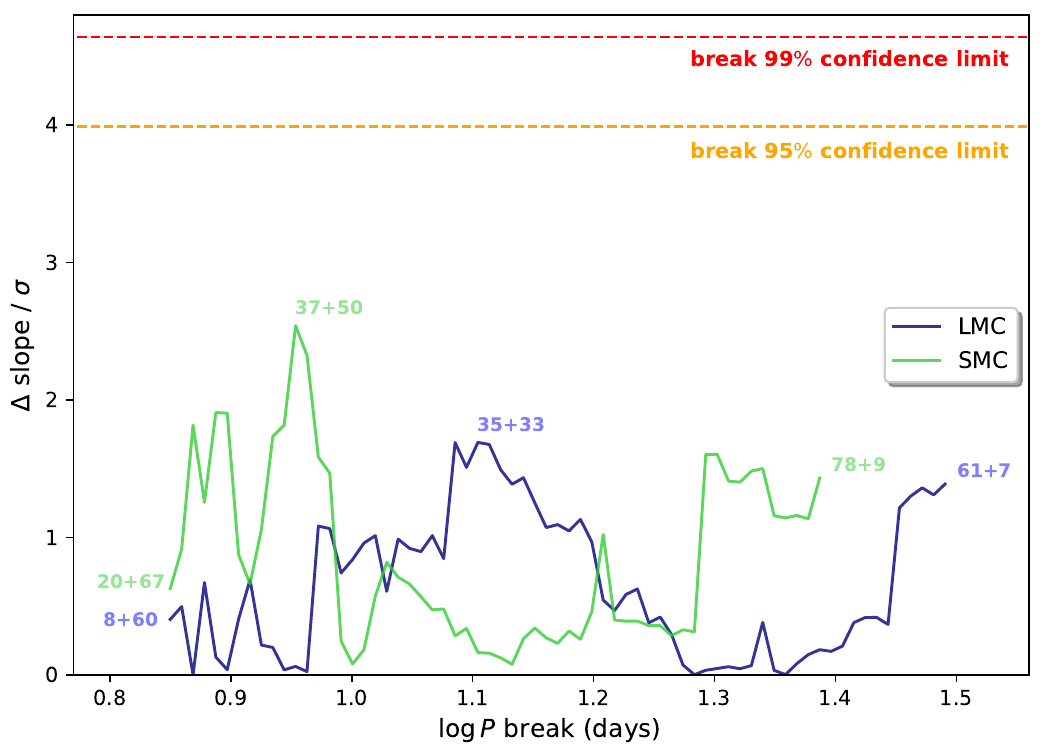}  
\caption{Slope difference between the long and short period regime divided by their errors in quadrature, for different values of the pivot period. The numbers $N+M$ give the number of Cepheids in the short (N) and long (M) period subsamples. \\
\label{fig:short_vs_long_period}}
\vspace{0.4cm}
\end{figure}

Here we investigate the presence of a possible discontinuity in the P--L slope \citep[see][]{Bhardwaj2016b}. In the present SMC sample and in the LMC sample from \citetalias{Riess2019}, we separate Cepheids into two subsamples, short vs long period Cepheids. For different values of the pivot period comprised between $\log P = 0.85$ and $\log P = 1.5$ (which leaves at least 10$\%$ of the initial sample on each side of the pivot period), we fit the P--L slope with a Monte Carlo algorithm taking into account the error bars and we compare the P--L slope in both regimes. Fig. \ref{fig:short_vs_long_period} shows the difference between the long- and short-period slopes divided by their errors in quadrature at different pivot periods. The LMC and SMC samples are represented in blue and green, and show a maximum slope difference of $1.8\sigma$ and $2.4\sigma$ respectively. 

However, we must account for the random chance of finding a break at any $\log P$. In order to estimate the likelihood of a break in the P--L relation , we create 10,000 random fake Cepheid samples (N=88) with the same P--L scatter as in the SMC ($\sigma=0.102$ mag). For each fake sample, we fit the P--L slope at short and long periods for a break located at different $\log P$ values. For these different $\log P$ breaks, we record the highest slope difference and we find that 51\% of the time,  the  slopes at short vs long periods disagree to higher than 2.4 times the combined error (i.e. higher than the value we found in the SMC). Fig. \ref{fig:short_vs_long_period} also shows the 99\% and 95\% confidence limits of this test. We conclude that it is not surprising to measure a change in the P--L slope which is 2.4 times the error as we find in the SMC because there is so much freedom in the search for a break. Regarding the position of the break, there is no preferred value but it occurs slightly more often at short ($\log P \sim 0.95$) and long ($\log P \sim 1.45$) periods, also where the samples are more limited. In conclusion, we find no evidence of a break in the P–L relation for our sample of LMC or SMC Cepheids.     \\



\newpage
\section{The SMC structure on smaller scales}

\begin{figure*}[h!]
\centering
\includegraphics[width=18.0cm]{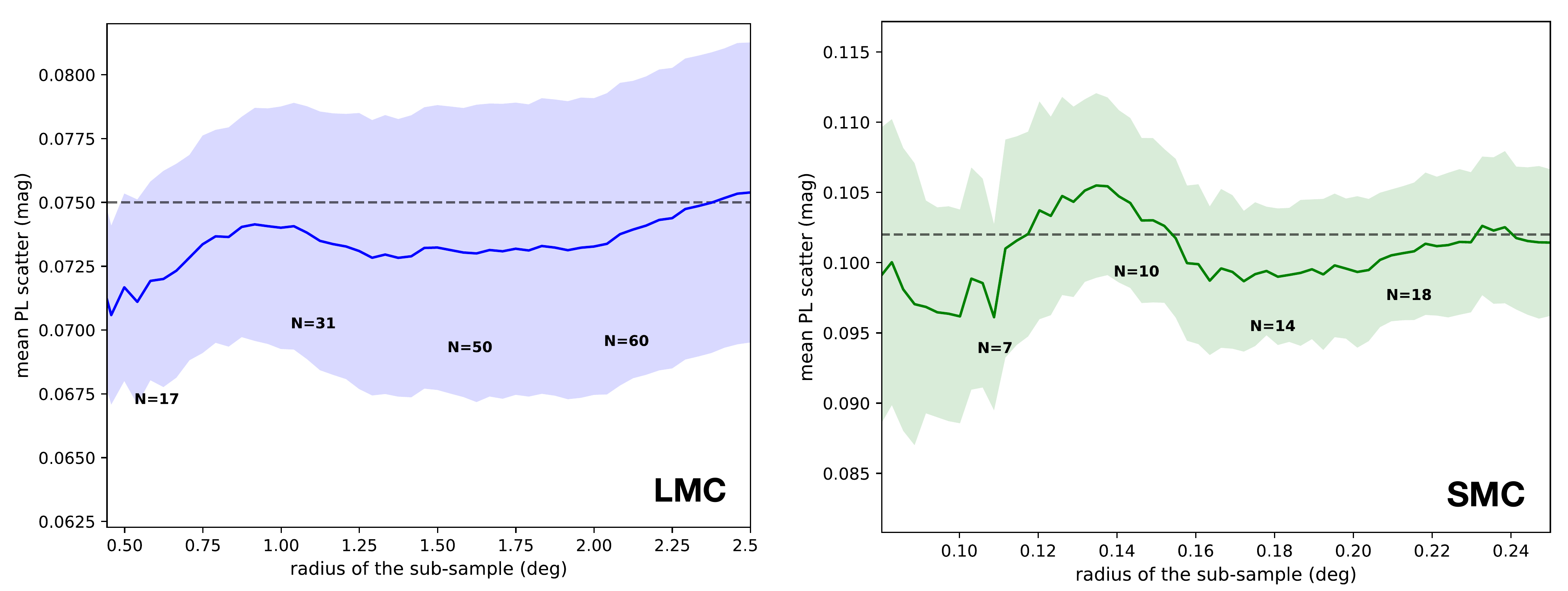}
\caption{Mean P--L scatter for Cepheid subsamples in different region sizes, in the LMC sample from \citetalias{Riess2019} (left) and our SMC sample (right). The mean number of Cepheids are shown in black for a few subsamples. The horizontal dashed line shows the scatter of the full Cepheid sample (in the LMC: $N=68$, $\sigma = 0.075 \, \rm mag$, in the SMC: $N=87$, $\sigma = 0.102 \, \rm mag$). \\
\label{fig:geometry_test_subsamples}}
\end{figure*}

Fig.~\ref{fig:geometry_test_subsamples} shows the effect of including progressively smaller regions in the LMC and SMC samples; although there is an apparent decrease in the dispersion in the SMC P--L relation for radii smaller than 0.12 deg, the effect is modest and of limited statistical significance.  We note that if the geometry of the SMC is akin to a bar, depth effects on scales comparable to the bar’s minor axis will remain even for very small regions near the center. \\


\bibliography{Breuval_2023.bib}{}
\bibliographystyle{aasjournal}

\end{document}